\documentclass[journal ]{new-aiaa}
\usepackage[utf8]{inputenc}
\usepackage{textcomp}

\usepackage{graphicx}
\usepackage{amsmath}
\usepackage[version=4]{mhchem}
\usepackage{siunitx}
\usepackage{longtable,tabularx}
\usepackage{color}

\title{Wall Ion Loss Reduction by Acceleration Zone Shifting in Anode-Layer Hall Thruster}

\author{Rei Kawashima \footnote{Assistant Professor, Department of Aeronautics and Astronautics; kawashima@al.t.u-tokyo.ac.jp. Member AIAA.},
Yushi Hamada\footnote{Graduate Student, Department of Advanced Energy. Student Member AIAA.},
Shu Kawabata\footnote{Graduate Student, Department of Aeronautics and Astronautics.},
Kimiya Komurasaki\footnote{Professor, Department of Aeronautics and Astronautics. Member AIAA.}}
\affil{The University of Tokyo, Tokyo, 113-8656, Japan}
\author{Hiroyuki Koizumi\footnote{Associate Professor, Department of Advanced Energy. Member AIAA.}}
\affil{The University of Tokyo, Kashiwa, 277-8561, Japan}

\begin{document}

\maketitle

\section*{Nomenclature}

{\renewcommand\arraystretch{1.0}
\noindent\begin{longtable*}{@{}l @{\quad=\quad} l@{}}
$B_r$  & radial magnetic flux density, T \\
$E_{\rm i,w}$ & ion energy colliding with wall, eV \\
$E_{\rm ion}$ & ionization potential, eV \\
$E_{\rm work}$ & work function of wall, eV \\
$e$ & elementary charge, C \\
$I_{\rm d}$   & discharge current, A \\
$I_{\rm g}$   & guard ring current, A \\
$i_{\rm g}$   & guard ring current density, A/m$^2$ \\
$i_{\rm i,w}$ & ion wall current density, A/m$^2$ \\
$M$ & atomic mass of wall material, kg \\
$P_{\rm i,w}$ & ion wall energy loss, W \\
$S$ & ion collecting surface area, m$^{2}$ \\
$V_{\rm d}$   & discharge voltage, V \\
$V_{\rm g}$   & guard ring voltage, V \\
$Y$           & sputter yield, atoms/ion\\
$Y_{\rm E}$   & energy factor of sputter yield, atoms/ion \\
$Y_{\rm v}$    & volume sputter yield, mm$^3$/C\\
$Y_{\theta}$   & angle factor of sputter yield, -- \\
$\gamma_{\rm e}$ & electron current factor, --\\
$\gamma_{\rm SEE}$ & secondary electron emission yield, --\\
$\varepsilon$ & wall erosion rate, \textmu m/h\\
$\rho$ & density of wall material, kg/m$^3$\\
\end{longtable*}}

\section{Introduction}
\lettrine{H}{all} thrusters are electric propulsion systems used in spacecraft.
They are equipped with an axisymmetric discharge channel with a crossed-field configuration, where radial magnetic and axial electric fields are applied.
Hall thrusters are available in two primary configuration types: a stationary plasma thruster (SPT) and a thruster with anode layer (TAL).
The distinguishing features of these configurations are discussed in the literature \cite{Zhurin:1999aa,Choueiri:2001aa}.
Compared with the SPT, the TAL demonstrates high-density plasma generation and a more compact thruster body with a short discharge channel.
Additionally, the TAL contains metallic channel walls biased at the cathode potential to repel the electrons approaching the walls, resulting in reduced electron energy losses to the channel walls.
The TAL has the potential for efficient plasma extraction; however, its superiority has not yet been demonstrated \cite{Mazouffre:2016aa}.

The optimal design and operation regime for TAL-type Hall thrusters with xenon and alternative propellants have been investigated \cite{Yamasaki:2019uo,Dukhopelnikov:2021wr}.
Hamada et al. developed a 5-kW TAL characterized by a magnetic field geometry where the peak of the magnetic flux density is located downstream of the channel exit \cite{HAMADA:2017aa}.
This thruster attained a maximum thrust efficiency of 0.60, which is comparable to the efficiencies of SPTs with the same input power level.
A 2-kW TAL was designed to incorporate discharge channel and magnetic field geometries similar to those of the 5-kW thruster \cite{Hamada:2021aa}.
The thruster size was reduced from that of the 5-kW thruster to maintain the flux density of the propellant xenon gas.
Probe diagnostics showed that the bulk plasma properties of the 5-kW and 2-kW TALs are similar.
Furthermore, these TAL thrusters exhibit primary acceleration zones outside the channel exit, owing to the abovementioned magnetic field design.
This plasma property is different from that of conventional TALs, in which the acceleration zone exists in the thin ``anode layer'' in front of the anode.

In a TAL with the acceleration zone shifted downstream, the plasma--wall interaction is critical for the energy efficiency and thruster lifetime.
If the potential of the bulk plasma inside the discharge channel is maintained near the anode potential, a fraction of the ions is inevitably attracted to the conducting channel walls.
The wall ion loss implies a loss of energy for ionization and acceleration.
In addition, the ions colliding with the wall display high energy because of the large potential drop between the bulk plasma and wall, leading to wall erosion due to ion bombardment.
Wall erosion is a significant concern in Hall thruster development because it is one of the primary limiting factors of thruster lifetime.
Several technologies have been proposed for wall erosion reduction in SPTs, including magnetic shielding \cite{Hofer:2014aa} and wall-less thrusters \cite{Vaudolon:2015te}.
Wall ion loss and erosion must be assessed for TALs with shifted acceleration zones.

The objective of this study was to measure the wall ion fluxes during TAL operation with a shifted acceleration zone and evaluate the wall erosion rate.
The ion wall flux was evaluated by measuring the metallic-channel wall current.
A previous study found that the wall current was dependent on the discharge voltage \cite{Hamada:2021aa}.
In this study, we further investigated the dependence of the ion wall flux on the magnetic configuration with a detailed analysis of the guard ring current.
A fast estimation method for the wall erosion rate using the guard ring current is proposed, and the erosion rate of the TAL with a shifted acceleration zone is evaluated.

\section{Experimental Setup}

\begin{figure}[hbt!]
    \centering
    \includegraphics[width=.3\textwidth]{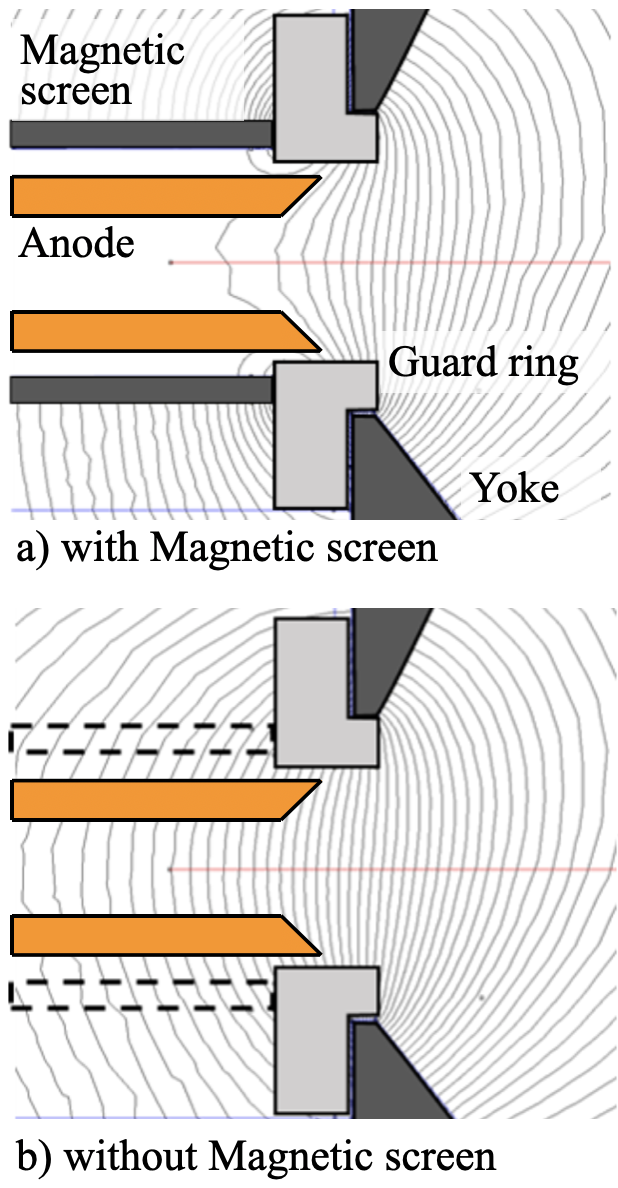}
    \caption{Magnetic field geometries in cases with and without magnetic screen.}
    \label{fig:magnetic}
\end{figure}

\begin{figure}[hbt!]
    \centering
    \includegraphics[width=.4\textwidth]{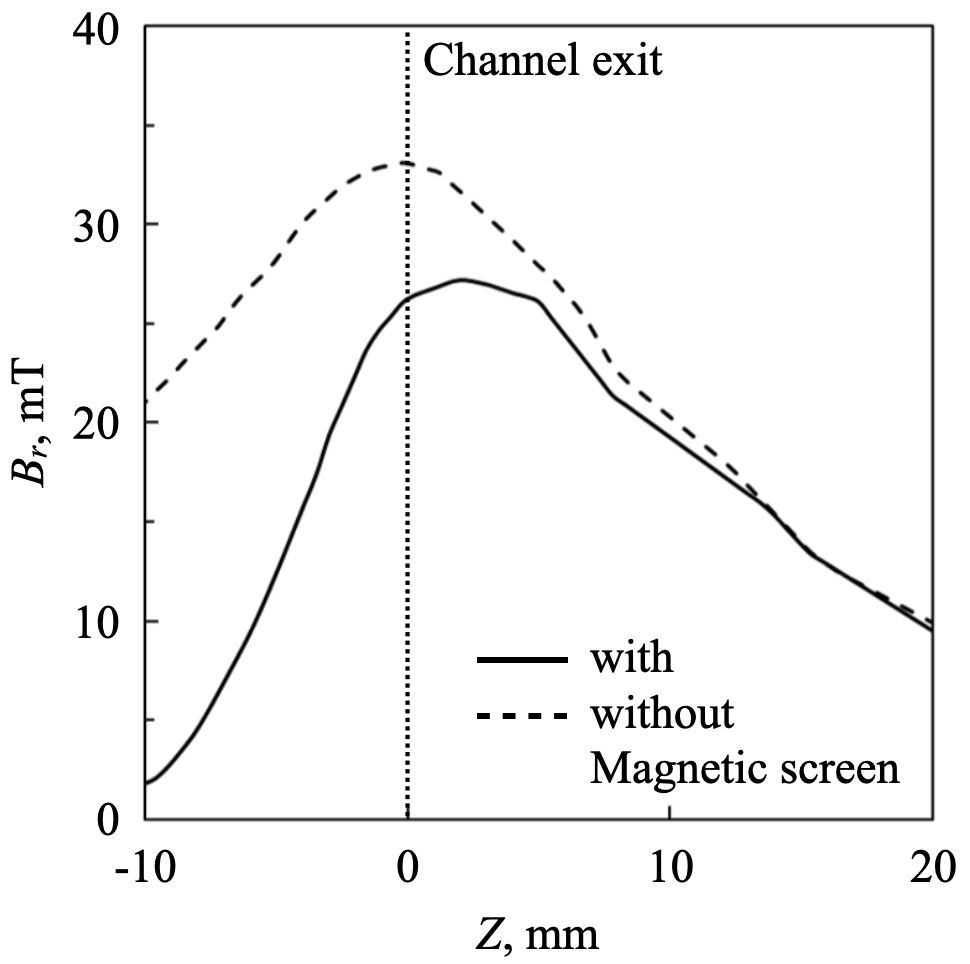}
    \caption{Magnetic flux density distributions along channel centerline.}
    \label{fig:br}
\end{figure}

The TAL thruster RAIJIN66 was used in the experiment \cite{Hamada:2021aa}.
The thruster contains a hollow copper anode, and the propellant xenon gas is supplied through the hollow anode.
The metallic channel wall is called the guard ring and is constructed from stainless steel.
A magnetic screen was installed in the magnetic circuit using recent SPT designs \cite{RezaIEPC2019}.
In the RAIJIN thrusters, the shifting of the acceleration zone downstream beyond the channel exit was achieved by using the magnetic screen effectively.
Two magnetic field geometries were tested (with and without the magnetic screen), as illustrated in Figs. \ref{fig:magnetic} and \ref{fig:br}.
When the magnetic screen was installed, the magnetic flux density upstream of the channel exit was trimmed, and thus, the region of strong magnetic confinement was concentrated downstream.
The position of the peak magnetic flux density was at $z = 0$ mm in the case without the magnetic screen and shifted to $z = 1.5$ mm in the case with the magnetic screen.
As displayed in Fig. \ref{fig:magnetic}, magnetic flux lines intersect the guard rings in both cases.
Thus, magnetic shielding was not implemented in this thruster.


\begin{figure}[hbt!]
    \centering
    \includegraphics[width=.4\textwidth]{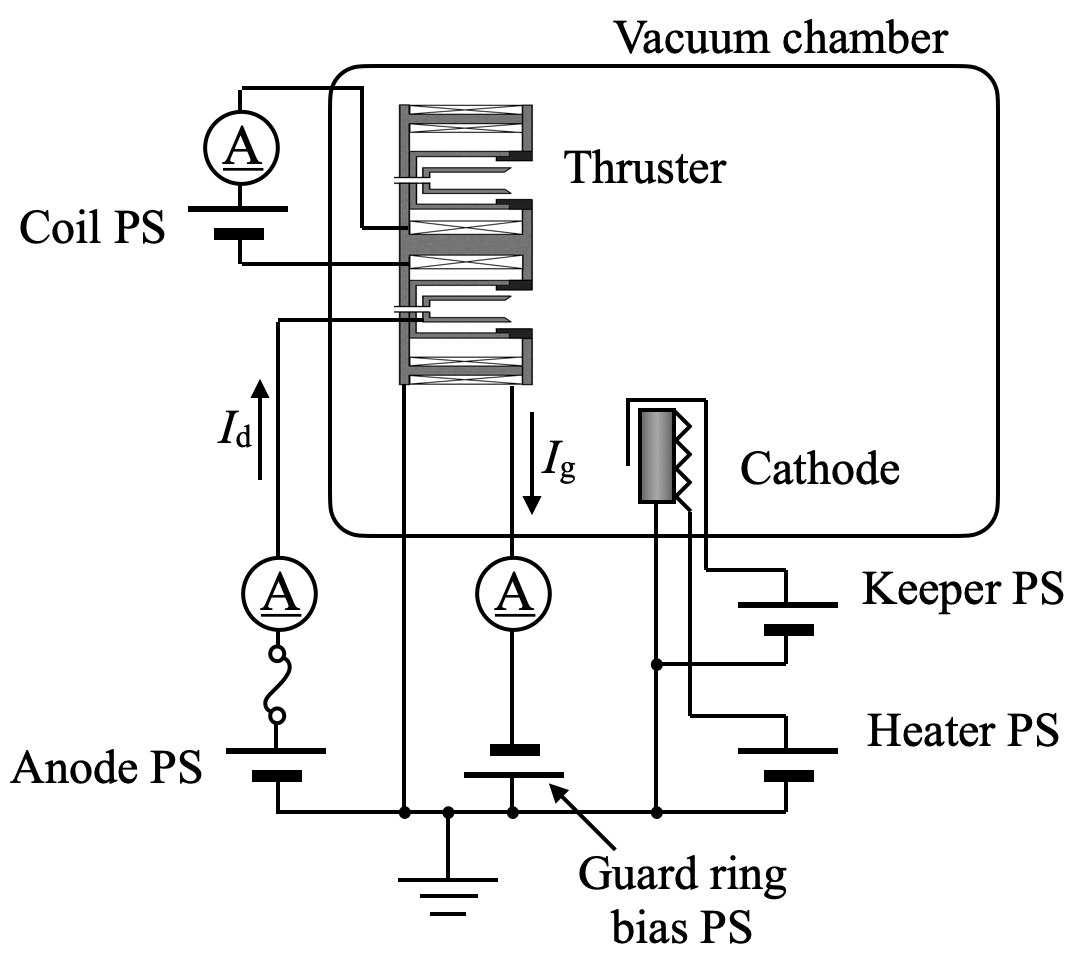}
    \caption{Circuit diagram of thruster operation and measurement.}
    \label{fig:circuit}
\end{figure}

The circuit diagram during the thruster operation is shown in Fig. \ref{fig:circuit}.
In nominal TAL operation, the guard rings and thruster body, including the front and back plates, are biased at the cathode potential.
The cathode body line was connected to the ground in this experiment for the stability of thruster operation.
In other words, the cathode-to-ground voltage was set to zero.
This would decrease the potential of the beam relative to the cathode and increase acceleration potential for ions compared with the case of floating cathode line.
Experimental studies have shown that the cathode-to-ground voltage is generally negative to the ground, and has a value of 3\% to 6\% of the discharge voltage \cite{Goebel:2012vb}.
We assumed that a variation of the acceleration potential over this range would not change the fundamental characteristics of wall ion losses.
In this experiment, the guard ring was isolated from the other thruster body parts, and $I_{\rm g}$ was defined as the current flowing into the inner and outer guard rings.
Additionally, a power supply was inserted to bias the guard rings to a potential lower than that of the cathode.
$I_{\rm d}$ and $I_{\rm g}$ were obtained by measuring the voltage drops across the 0.5-$\Omega$ shunt resistors.
Data acquisition was implemented by differential probes and an oscilloscope.

\begin{figure}[hbt!]
    \centering
    \includegraphics[width=.4\textwidth]{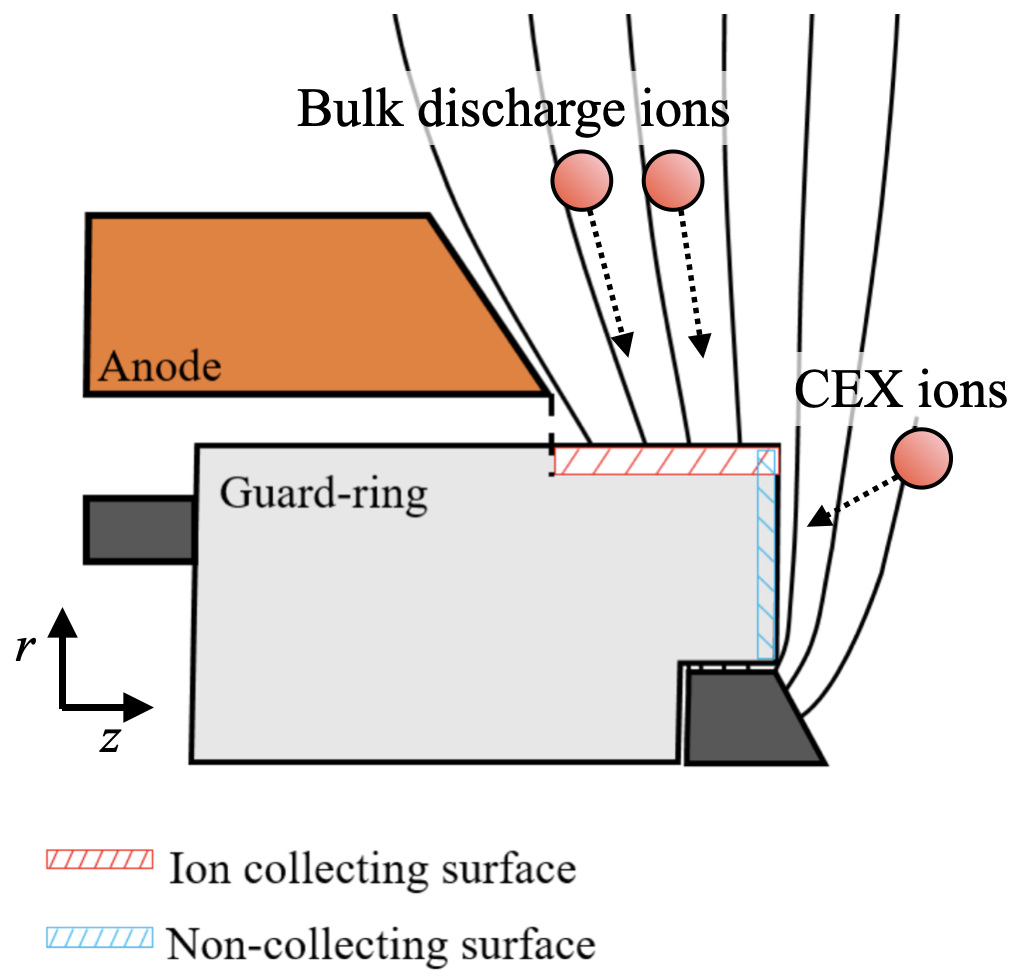}
    \caption{Close-up view of the guard ring.}
    \label{fig:collect}
\end{figure}

To calculate the guard ring current density $i_{\rm g}$, the ion-collecting surface area $S$ must be defined.
A close-up view of the guard ring and magnetic lines is shown in Fig. \ref{fig:collect}.
The guard rings made from type 304 stainless steel are in contact with the plasma at the surfaces in the r- and z-directions.
The bulk discharge ions generated in the discharge channel flow into the guard ring from the surface normal to the r-direction.
The surface normal to the z-direction predominantly collect the ions generated in the charge-exchange collision (CEX) process.
It is assumed that the contribution of the CEX ions to the guard ring current is small compared to that of the bulk discharge ions.
Therefore, the ion-collecting area is defined as the surface normal to the r-direction, with an axial length of 2.1 mm (distance between the hollow anode tip and guard ring end) and outer and inner diameters of 66 mm and 54.1 mm, respectively.
Note that neglecting the z-direction surfaces results in a slight overestimation of $i_{\rm g}$ (and thus of the wall erosion) because the measured $I_{\rm g}$ includes both the bulk discharge and CEX ions.

A vacuum chamber with a diameter of 2 m and length of 3 m was used in the experiment.
During the experiment, the background pressure was maintained below 6.6$\times$10$^{-3}$ Pa.
The hollow cathode HCN-252 (Veeco Instruments Inc.) was used as the electron emitter, with a xenon flow rate of 0.29 mg/s.
The background xenon particles inevitably affect the thruster performance and discharge plasma characteristics.
The background pressure effects were investigated for 1.35-kW SPT-100, and it was reported that the backpressure of 6.7$\times$10$^{-3}$ Pa caused a non-negligible performance change such as 4\% increase in thrust, because a fraction of the background particles worked as propellant \cite{Diamant:2014wu}.
In addition, it was reported that the acceleration zone moved upstream by a few millimeters owing to the backpressure \cite{MacDonald-Tenenbaum:2019wt}.
In this experiment, it was expected that these background pressure effects would increase the measured $i_{\rm g}$.
Thus, the analysis given in this study may yield a slightly overestimated wall erosion rate.

The anode mass flow rate of the xenon propellant was set to 3.35 mg/s, and the discharge voltage was set to either 200 V or 300 V.
Under these conditions, two experiments were carried out: 1) the current--voltage (I-V) characteristics of the guard rings were obtained to examine how much electron current was included in the guard ring current, and 2) $i_{\rm g}$ was measured for different magnetic configurations with various magnetic field strengths.
The systematic errors of the measurement system were estimated to be $\pm 3.7$\% for $I_{\rm d}$ and $I_{\rm g}$, and $\pm 10.2$\% for $i_{\rm g}$ based on the errors in the shunt resistors, differential probes, oscilloscope, and machining of guard rings.

\section{Results and Discussion}

\subsection{I-V Characteristics of Guard Ring}
\begin{figure}[hbt!]%
    \centering
    \includegraphics[width=.4\textwidth]{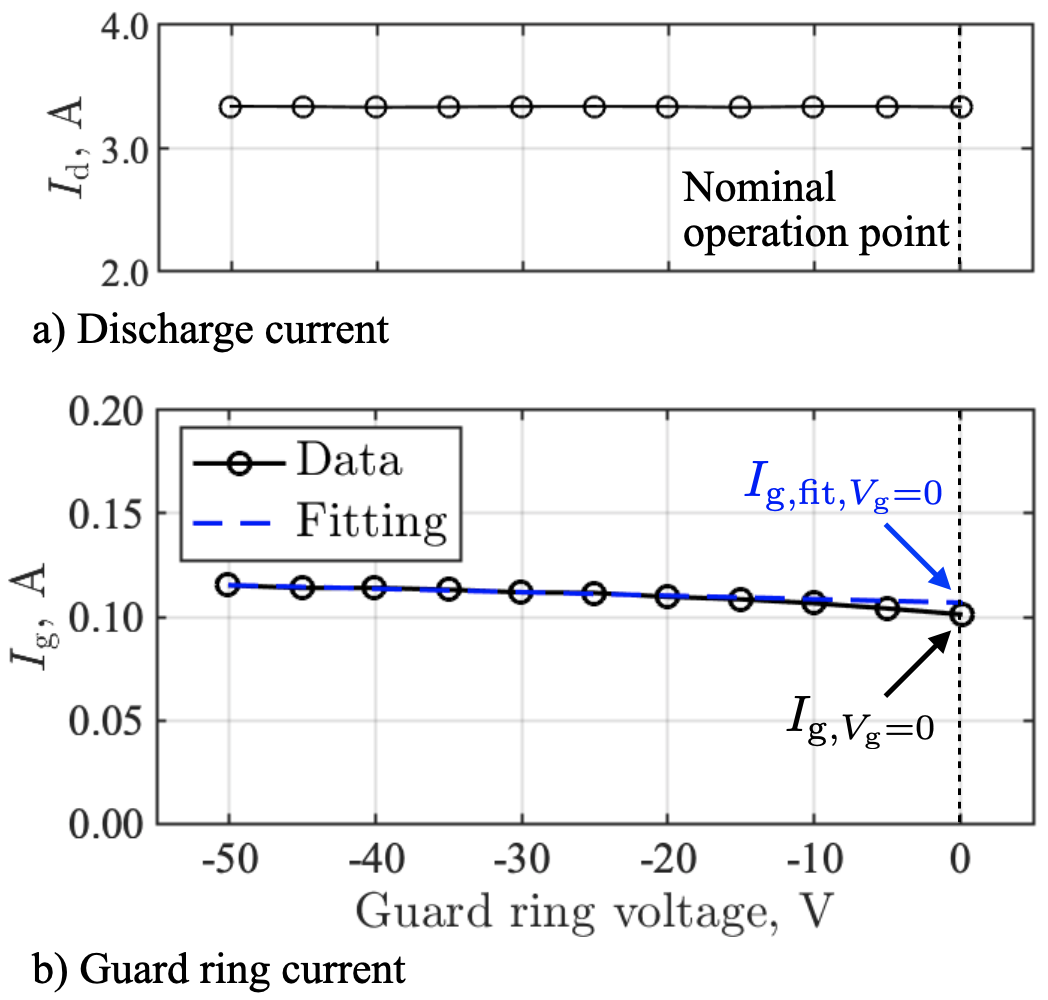}
    \caption{Discharge and guard-ring currents with varied guard-ring voltages.}
    \label{fig:bias}
\end{figure}

The dependence of $I_{\rm d}$ and $I_{\rm g}$ on the guard ring bias voltage $V_{\rm g}$ is illustrated in Fig. \ref{fig:bias}.
In this experiment, the discharge voltage and peak magnetic flux density were fixed at 200 V and 23 mT, respectively.
The typical errors for $I_{\rm d}$ and $I_{\rm g}$ are $\pm$0.13 A and $\pm$4.5 mA, respectively.
To examine if the guard ring sheath was in the ion saturation regime, $V_{\rm g}$ was varied over a range of $-50$ V$<V_{\rm g}<$ 0 V.
During the sweep of $V_{\rm g}$, $I_{\rm d}$ was maintained close to 3.34 A, and hence, the effect of the guard ring voltage on the bulk discharge plasma was insignificant.
$I_{\rm g}$ exhibited a slight dependence on $V_{\rm g}$.
Linear fitting was conducted using the data for $-50$ V$<V_{\rm g}< - 20$ V, and the ion saturation current was estimated by extrapolating the fitting line to $V_{\rm g}$ at 0 V.
The dimensionless factor $\gamma_{\rm e}$ was used to indicate the electron current included in the guard ring current, which is defined as $1+\gamma_{\rm e}=\left(I_{\rm g,fit}/I_{\rm g}\right)_{V_{\rm g}=0}$.
$\gamma_{\rm e}$ of 0.044 was obtained from the measurements.
The small $\gamma_{\rm e}$ indicates that the sheaths on the guard rings are in the ion saturation regime and that the guard ring current can be used to evaluate the wall ion fluxes.
A $\gamma_{\rm e}$ value of 0.044 was used in the wall ion current analysis described in Sec. \ref{sec:erosion}.

\subsection{Guard Ring Current Measurement Result}

\begin{figure}[t!]
    \centering
    \includegraphics[width=.4\textwidth]{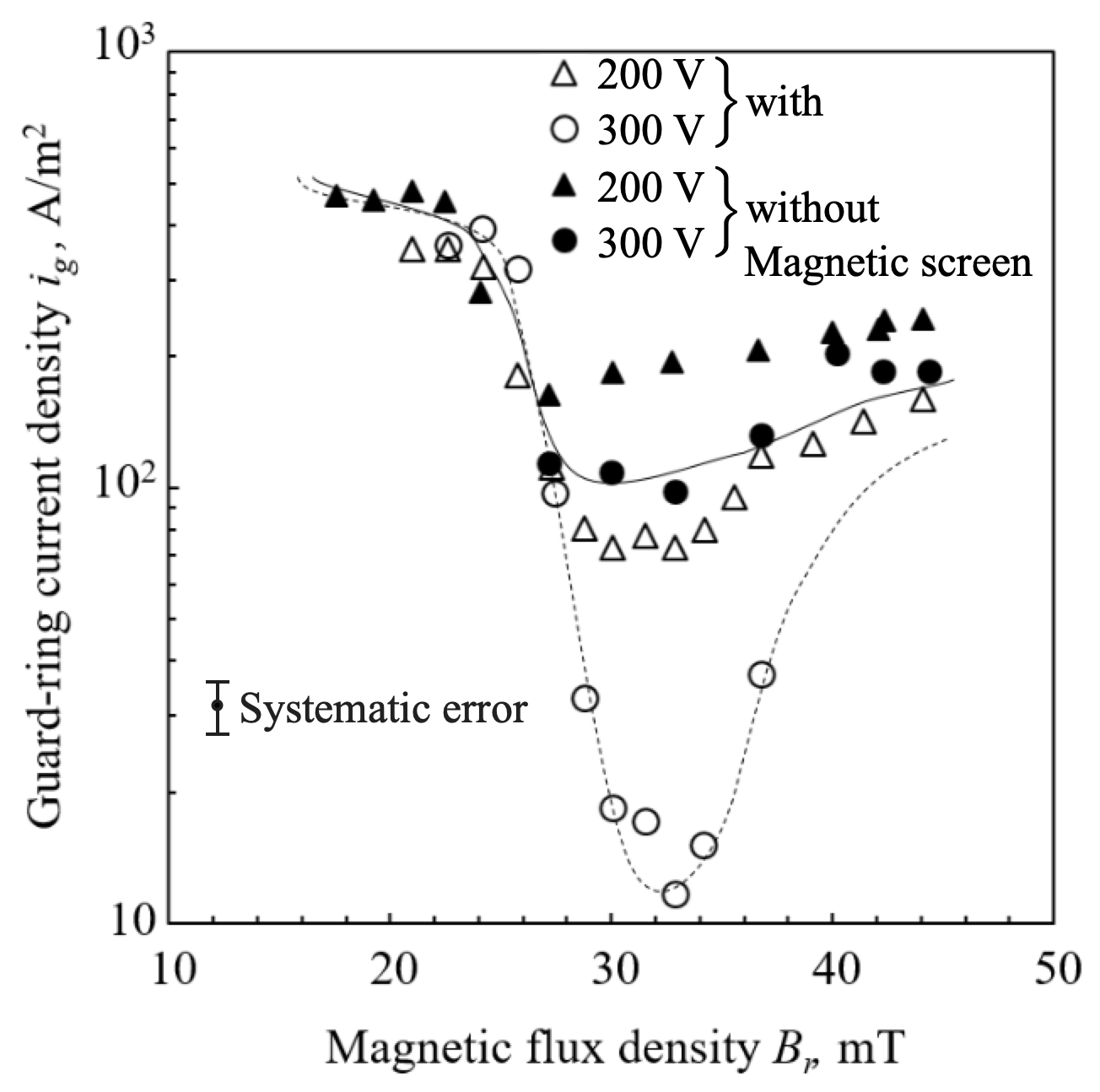}
    \caption{Guard ring current densities for various magnetic flux densities.}
    \label{fig:ig}
\end{figure}

Figure \ref{fig:ig} displays the guard ring current density corresponding to various magnetic flux densities for the cases with and without the magnetic screen.
$i_{\rm g}$ is typically dependent on $B_{r}$, indicating that an optimal magnetic flux density exists in which the guard ring current is minimized.
At $V_{\rm d}=300$ V and $B_r=32$ mT, a minimum $i_{\rm g}$ of 12 A/m$^{2}$ was obtained during operation with the magnetic screen, whereas a minimum $i_{\rm g}$ of 98 A/m$^{2}$ was obtained when the magnetic screen was not used.

In TALs, the ions colliding with the guard rings induce secondary electron emission (SEE).
The secondary electrons emitted from the guard ring surface increase the measured current.
The wall ion current density was obtained from the guard ring current density as $i_{\rm i,w}=\left(1+\gamma_{\rm e}\right)/\left(1+\gamma_{\rm SEE}\right)\cdot i_{\rm g}$, where $\gamma_{\rm SEE}$ is the SEE yield.
When the acceleration potential of impact ion is in the range of 10--1000 V, the process of SEE is realized by Auger emission \cite{LiebermanTextbook}.
The ion recombination on the wall surface is a three-body recombination process where two electrons are involved.
When an electron enters the ground state of the impact ion, the energy corresponding to $E_{\rm ion}$ is released.
Because the first electron is emitted from the surface, the energy of $E_{\rm work}$ is consumed from the ionization potential.
The excess energy, $E_{\rm ion}-E_{\rm work}$ is transferred to the secondary electron.
The secondary electron must gain an energy greater than $E_{\rm work}$ to move away from the surface.
Thus, the condition for SEE is $E_{\rm ion}-2E_{\rm work}\geq 0$.
In this study, we used an empirical expression to estimate the SEE yield: $\gamma_{\rm SEE}\approx 0.016\left(E_{\rm ion}-2E_{\rm work}\right)=0.055$ \cite{LiebermanTextbook}.
Here, $E_{\rm ion}=12.13$ eV is the ionization potential of xenon, and $E_{\rm work}=4.34$ eV \cite{Wilson:1966vw} is the work function of type 304 stainless steel.
Using these relationships, the wall ion current density can be estimated from the guard-ring current density as $i_{\rm i,w}=0.99i_{\rm g}$.
This conversion coefficient was assumed to be constant for the cases of different magnetic configurations.

The estimated $i_{\rm i,w}$ is 97 A/m$^{2}$ for the case without the magnetic screen and 12 A/m$^{2}$ for the case with the magnetic screen.
$i_{\rm i,w}$ is reduced because the acceleration and ionization zones move downstream when the magnetic screen is used, and the ions are exhausted without colliding with the guard rings.
Therefore, the wall ion loss in a TAL can be reduced using a magnetic screen and the optimal magnetic flux density.

\subsection{Wall Erosion Analysis}
\label{sec:erosion}

The wall erosion rate was estimated based on the measured wall ion current and sputter yield.
This estimation method allows for a quick assessment of the wall erosion rate in TALs.
The sputter yield $Y$ (atoms/ion) was modeled as $Y=Y_E\times Y_\theta$, where $Y_E$ and $Y_\theta$ represent the functions of the dependences of the incident ion energy and angle, respectively.
Emissive probe measurements of the TAL thruster demonstrated that the plasma potential was close to the anode voltage inside the discharge channel \cite{Hamada:2021aa}.
Thus, the acceleration potential of ions colliding with the cathode-biased wall was assumed to be as high as $V_{\rm d}$.
A $Y_{\rm E}$ of 0.55 was assumed for 300-eV xenon ion bombardment on stainless steel walls based on experimental data \cite{Andersen:1980up}.
In this analysis, the wall erosion rate was estimated for the graphite wall material.
Carbon graphite exhibits durability against ion bombardment, and a $Y_{\rm E}$ of 0.08 was assumed for a 300-eV xenon impact based on published experimental data \cite{Kenmotsu:2009vc,Boyd:2001wm}.
$Y_\theta$ is a normalized quantity relative to the case of normal incidence.
Several studies have shown that $Y_\theta$ generally increases as the incident angle of the impact ion increases, and $Y_\theta$ has a peak when the incident angle is around 45--80 deg \cite{Yamamura:1984uf,Ranjan:2016wu}.
In the thruster operation with the magnetic screens, the axial acceleration inside the channel was negligible compared with the radial acceleration due to the potential drop in $V_{\rm d}$ toward the guard rings, concerning the ions colliding with the walls.
Therefore, the normal incidence case was assumed as $Y_\theta=1$.
In the case without the magnetic screens, there would be a possibility that the acceleration zone shifted upstream and the ion incidences were in oblique.
However, because the primary purpose of the present analysis was to analyze the erosion rate of the thruster with the magnetic screens, we assumed $Y_\theta$ of unity for both magnetic configurations.
The volumetric sputter yield is calculated as $Y_{\rm V}=YM/\left(e\rho\right)$, where $M$ and $\rho$ are the mean atomic weight and mass density of the wall material.
The wall erosion rate can be estimated as $\varepsilon=Y_{\rm V}i_{\rm i,w}=Y_E Y_\theta M/\left(e\rho\right)\cdot i_{\rm i,w}$.

The wall ion energy loss $P_{\rm i,w}$ and erosion rate $\varepsilon$ estimated for RAIJIN66 are listed in Table \ref{tab:erosion}.
As a reference, the SPT data in the unshielded and magnetically shielded configurations are also shown \cite{Hofer:2014aa}.
Here, the power corresponding to the wall ion energy loss was estimated as $P_{\rm i,w}=i_{\rm i,w}S\left(E_{\rm ion}+E_{\rm i,w}\right)$.
In RAIJIN66, by installing a magnetic screen, the wall ion loss was reduced by approximately one-eighth.
The wall ion energy loss power for the case with the magnetic screen was reduced to 2.6 W, which was less than 0.3\% of the 1 kW anode input power.
The wall erosion rate was also reduced to 1.68 \textmu m/h for the stainless steel walls.
Furthermore, if graphite is used as the guard ring material, $\varepsilon$ can be expected to be 0.18 \textmu m/h.
This value implies that the wall recession is expected to be 1.8 mm for a 10000 h operation.
The guard rings of RAIJIN66 are 2.0 mm thick, and therefore endure the erosion.
It is concluded that a thruster lifetime exceeding 10,000 h is possible in TALs with a downstream-shifted acceleration zone.

\begin{table*}[t]
\caption{\label{tab:erosion} Energy loss and wall erosion rate estimation for RAIJIN66 and H6 \cite{Hofer:2014aa}.}
\centering
\begin{tabular}{p{18mm}p{10mm}p{30mm}ccccc}
    \hline
    Thruster & Wall & Configuration & $i_{\rm i,w}$, A m$^{-2}$ & $E_{\rm i,w}$, eV &$Y_{\rm v}, \times10^{-3}$ mm$^{3}$ C$^{-1}$ &$P_{\rm i,w}$, W &  $\varepsilon$, \textmu m/h \\
    \hline
    RAIJIN66  & SUS  & without mag. screen   & 97   & 300 & 39 & 22 & 13.7 \\
    RAIJIN66   & SUS & with mag. screen & 12   & 300 &39 &2.6 & 1.68 \\
    RAIJIN66   &  C & with mag. screen & 12   & 300 &4.3 &2.6 & 0.18 \\
    H6   &  BN &unshielded   & 210 &   143  & 28 &--& 21.1 \\
    H6   & BN & magnetic shielding & 40 &    36.1 & 0.14 &--& 0.02\\
    \hline
\end{tabular}
\end{table*}

    Although the wall erosion of RAIJIN66 with a carbon graphite wall was expected to be minor, the wall erosion rate was greater than that in the magnetically shielded SPT by one order of magnitude.
To further extend the TAL thruster lifetime, chamfering or curving of the guard ring, such as with the magnetically shielded thrusters, may be an effective method.
In addition, we identified the optimal operation condition in terms of the wall ion loss and wall erosion.
It was found that this optimal operation condition deviated from the point of the highest thruster performance.
The anode efficiency of RAIJIN66 was obtained for various operating conditions \cite{Hamada:2021aa}.
For the case with discharge voltage and anode mass flow rate of 300 V and 3.35 mg/s, $B_r$ of 70 mT yielded higher anode efficiency compared with the case of $B_r=32$ mT.
This means that there is a trade-off between the wall erosion and anode efficiency in the present thruster design.
Further exploration of the optimal operating condition and optimization of thruster design are envisaged for a future study.

\section{Conclusion}
The wall ion loss and erosion rate were investigated for a TAL-type Hall thruster characterized by an acceleration zone shifted downstream.
The guard rings were electrically separated from the thruster body, and the guard ring current was measured independently.
The I-V characteristics of the guard ring indicated that the sheath in front of the guard rings was in the ion saturation regime and that the guard ring current can be used to indicate the wall ion flux, enabling a fast evaluation of the wall erosion rate.
The wall ion current, wall ion energy loss, and erosion rate were estimated based on the measured guard ring current.

The results demonstrate that the wall ion loss can be significantly reduced using a magnetic screen and the optimal magnetic flux density.
The wall ion energy loss was less than 0.3\% of the anode input power.
The wall erosion rate was estimated to be 1.68 \textmu m/h and 0.18 \textmu m/h when using stainless steel and carbon walls, respectively.
If carbon material is used, a thruster lifetime exceeding 10000 h is likely.
It has been demonstrated that improved thruster lifetime can be achieved in the TAL, with the acceleration zone shifted downstream.

\section*{Funding Sources}

This work was supported by JSPS KAKENHI Grant Number JP20H02346.

\bibliography{reference}

\end{document}